\documentclass[%
 aip,
 amsmath,amssymb,
 reprint,%
]{revtex4-1}
\usepackage{algorithm}        
\usepackage{algorithmicx}
\usepackage{graphicx}
\usepackage{dcolumn}
\usepackage{bm}
\usepackage{booktabs}
\usepackage{threeparttable}
\usepackage[utf8]{inputenc}
\usepackage[T1]{fontenc}
\usepackage{mathptmx}
\usepackage{etoolbox}
\usepackage{array} 
\usepackage{subcaption}
\usepackage{siunitx}
\usepackage{orcidlink}
\usepackage[abs]{overpic}
\usepackage{xcolor}
\usepackage{multirow}
\usepackage{makecell}       
\usepackage[table]{xcolor}
\makeatletter
\def\@email#1#2{%
 \endgroup
 \patchcmd{\titleblock@produce}
  {\frontmatter@RRAPformat}
  {\frontmatter@RRAPformat{\produce@RRAP{*#1\href{mailto:#2}{#2}}}\frontmatter@RRAPformat}
  {}{}
}%
\makeatother
\begin{document}

\preprint{AIP/123-QED}

\title{High-Fidelity Reconstruction of Charge Boundary Layers and Sharp Interfaces in Electro-Thermal-Convective Flows via Residual-Attention PINNs} 



\author{Baitong Zhou}
\thanks{These authors contributed equally to this work.}
\affiliation{Nanophotonics and Biophotonics Key Laboratory of Jilin Province, School of Physics, Changchun University of Science and Technology, Changchun 130022, P.R. China}

\author{Ze Tao}
\thanks{These authors contributed equally to this work.}
\affiliation{Nanophotonics and Biophotonics Key Laboratory of Jilin Province, School of Physics, Changchun University of Science and Technology, Changchun 130022, P.R. China}

\author{Ke Xu}
\affiliation{Nanophotonics and Biophotonics Key Laboratory of Jilin Province, School of Physics, Changchun University of Science and Technology, Changchun 130022, P.R. China}

\author{Fujun Liu}
 \altaffiliation{Author to whom correspondence should be addressed: fjliu@cust.edu.cn}
\affiliation{Nanophotonics and Biophotonics Key Laboratory of Jilin Province, School of Physics, Changchun University of Science and Technology, Changchun 130022, P.R. China}

\author{Xuan Fang}
 \altaffiliation{Author to whom correspondence should be addressed: fangx@cust.edu.cn}
\affiliation{Nanophotonics and Biophotonics Key Laboratory of Jilin Province, School of Physics, Changchun University of Science and Technology, Changchun 130022, P.R. China}
\affiliation{School of Physics, Changchun University of Science and Technology, Changchun 130022, P.R. China}

\begin{abstract}
Accurate reconstruction of localized extreme structures remains a critical bottleneck in the physics-informed modeling of electro-thermal-convective flows. Although conventional physics-informed neural networks effectively capture smooth global dynamics, they frequently suffer from numerical diffusion and distortion when attempting to resolve sharp charge boundary layers or abrupt multiphase interfaces. To address these limitations, we propose a Residual-Attention Physics-Informed Neural Network (RA-PINN) that embeds gated attention modulation within a residual feature framework to adaptively enhance local sensitivity to steep physical gradients. The proposed architecture is rigorously evaluated against standard and recurrent network baselines using canonical electrohydrodynamic scenarios, encompassing near-electrode exponential boundary layers and sharply concentrated charge fields. Quantitative analyses demonstrate that the RA-PINN significantly reduces localized errors and faithfully preserves critical interface topologies without compromising the global consistency dictated by the coupled governing equations. Ultimately, this methodology establishes a highly robust predictive framework for resolving complex interfacial and boundary layer phenomena in advanced fluid dynamics applications.
\end{abstract}

\pacs{}

\maketitle 

\section{Introduction}
\label{sec:introduction}

Complex coupled multi-physics fields are widely encountered in advanced fluid dynamics applications, particularly within electro-thermal-convective systems, microscale transport, and electrohydrodynamics. In these environments, the fundamental physical states are often governed by strongly coupled variables such as velocity, pressure, temperature, and electric potential. A critical characteristic of these electro-thermal-flow systems is the emergence of highly localized extreme structures, including near-wall boundary-layer amplification, abrupt multiphase interfaces, and compact charge concentrations. These localized patterns fundamentally dictate transport intensity, interfacial stability, and thermal distribution, making them far more critical to system dynamics than the smooth global background flow. Consequently, developing robust predictive models capable of accurately reconstructing these steep spatial gradients and interfacial topologies remains a formidable challenge in computational fluid dynamics.

Traditional numerical methods provide high fidelity for coupled-field problems but incur prohibitive computational costs, often requiring highly refined adaptive meshes or specialized interface-capturing algorithms to resolve sharp gradients. Physics-informed neural networks (PINNs) have emerged as a compelling alternative, unifying governing partial differential equations, physical boundary conditions, and observational data into a cohesive continuous learning framework \cite{raissi2019pinn,karniadakis2021piml}. The efficacy of PINNs has been extensively validated across broad domains of fluid mechanics, heat transfer, and complex multi-physics modeling \cite{cai2021heattransfer,jagtap2020xpinns,yu2022gpinn,chen2025pfpinns}. This capability has driven the recent expansion of physics-informed methodologies toward data-efficient surrogate construction and advanced predictive modeling in complex engineering systems \cite{jeong2024data,wu2024reviewpiml,shi2024pinnc,chen2025epinn}. Furthermore, targeted PINN architectures have been recently proposed to address difficult dynamic fluid interfaces and nonlinear transient transport phenomena \cite{xing2025modeling,tao2025lnn,tao2025analytical}. Nevertheless, when applied to fields containing narrow boundary layers or sharp multiphase transitions, standard PINNs persistently suffer from optimization pathologies, spectral bias, and severe numerical diffusion, ultimately failing to preserve local structural fidelity \cite{wang2021gradient,wang2022ntk,sarma2024ipinns,tseng2023cusp}.

Enhancing the foundational network architecture provides a systematic pathway to overcome these representational limitations in physics-informed fluid modeling. Residual learning paradigms have been shown to stabilize feature propagation across deep computational graphs and mitigate optimization degradation \cite{he2016resnet}. Simultaneously, attention mechanisms offer the ability to adaptively amplify spatially informative patterns, thereby sensitizing the model to structurally critical regions with steep gradients \cite{vaswani2017attention}. Recurrent architectures such as Long Short-Term Memory (LSTM) have also been integrated into physics-informed frameworks to improve hidden-state spatial propagation and nonlinear feature interaction \cite{hochreiter1997lstm,wang2025pirn}. Building upon these representational advancements, this study develops a Residual-Attention Physics-Informed Neural Network (RA-PINN) specifically engineered to resolve the numerical diffusion inherent in capturing localized extreme fluid structures.

To systematically evaluate the physical representational capability of the proposed RA-PINN, three canonical electrohydrodynamic benchmark scenarios are constructed. The first scenario involves an exponential electric-charge boundary layer near an electrode, rigorously testing the model capacity to preserve strongly amplified near-wall behavior under transverse modulation. The second scenario features a ring-shaped abrupt interface induced by a cylindrical electrode, designed to evaluate the preservation of annular multiphase transitions without yielding to excessive radial smoothing. The third scenario introduces a circular sharp-interface field enclosing a compact local charge concentration, challenging the architecture to simultaneously reconstruct a highly localized charged core alongside continuous surrounding fluid fields. Under a unified physics-constrained training environment, the RA-PINN is comprehensively benchmarked against both standard pure PINN and LSTM-PINN architectures to validate its superior structural similarity and minimal local maximum error.

The remainder of this paper is structured to detail this systematic investigation. Section \ref{sec:governing} introduces the coupled governing equations and the methodology for constructing the analytic fluid benchmarks. Section \ref{sec:method} outlines the proposed RA-PINN architecture, detailing the residual-attention mechanisms and the unified physics-informed training workflow. Section \ref{sec:benchmark} specifies the physical parameters and experimental settings for the three electrohydrodynamic scenarios. Section \ref{sec:results} presents a comprehensive comparative analysis of the numerical predictions and field reconstructions. Finally, Section \ref{sec:engineering} discusses the physical implications of the proposed model for resolving complex electro-thermoconvective flows, followed by concluding remarks in Section \ref{sec:conclusion}.

\section{Electro-Thermoconvective Governing Equations and Physical Constraints}
\label{sec:governing}

\subsection{Coupled Electrohydrodynamic and Thermal Transport Equations}

Consider a two-dimensional electro-thermal-convective flow within a computational domain $\Omega\subset\mathbb{R}^2$ parameterized by spatial coordinates $\mathbf{x}=(x,y)$. The fundamental state of this physical system is completely described by the coupled continuous field vector 
\begin{equation}
\mathbf{U}(x,y)=[u(x,y),v(x,y),p(x,y),T(x,y),\phi(x,y)],
\end{equation}
where $u$ and $v$ represent the orthogonal fluid velocity components, $p$ denotes the kinematic pressure, $T$ signifies the continuous temperature field, and $\phi$ is the electric potential. The spatial distribution and interaction of these variables are strictly governed by the incompressible Navier-Stokes equations, the convection-diffusion energy equation, and a Poisson-type equation for the electric field. Formally, this coupled system is expressed as
\begin{equation}
\nabla\cdot\mathbf{u}=0,
\end{equation}
\begin{equation}
\mathbf{u}\cdot\nabla u+\frac{\partial p}{\partial x}-\nu\nabla^2 u-f_x(\phi,T)=s_u(x,y),
\end{equation}
\begin{equation}
\mathbf{u}\cdot\nabla v+\frac{\partial p}{\partial y}-\nu\nabla^2 v-f_y(\phi,T)=s_v(x,y),
\end{equation}
\begin{equation}
\mathbf{u}\cdot\nabla T-\alpha\nabla^2 T-q_T(\phi)=s_T(x,y),
\end{equation}
\begin{equation}
\nabla^2\phi-\lambda_{\phi}\phi=s_{\phi}(x,y).
\end{equation}
Within these governing equations, $\nu$ represents the kinematic viscosity of the fluid and $\alpha$ defines its thermal diffusivity. The highly nonlinear coupling between the hydrodynamic, thermal, and electric fields is driven by the internal forcing terms $f_x(\phi,T)$ and $f_y(\phi,T)$, which encapsulate local electrohydrodynamic body forces and thermally induced buoyancy effects. Concurrently, the term $q_T(\phi)$ accounts for thermal energy generation, such as localized Joule heating induced by the electric potential, while $\lambda_{\phi}$ dictates the specific reaction or screening coefficient within the electrostatic field. The spatially dependent external forcing terms $s_u$, $s_v$, $s_T$, and $s_{\phi}$ are analytically prescribed to ensure strict energy and momentum conservation across specific benchmark configurations.

\subsection{Method of Manufactured Solutions and Flow Configurations}

To rigorously isolate the representational capability of the neural network architectures from classical numerical discretization errors and experimental uncertainties, this study employs the Method of Manufactured Solutions. Under this robust mathematical paradigm, exactly known analytical fields are prescribed a priori for the entire computational domain. Let this exact, physically consistent solution be denoted by
\begin{equation}
\mathbf{U}^{\ast}(x,y)=[u^{\ast}(x,y),v^{\ast}(x,y),p^{\ast}(x,y),T^{\ast}(x,y),\phi^{\ast}(x,y)].
\end{equation}
By substituting these exact profiles directly into the governing electro-thermoconvective system, the corresponding residual source terms are analytically derived, ensuring that $\mathbf{U}^{\ast}$ satisfies the coupled partial differential equations pointwise. This rigorous strategy guarantees the availability of continuous, exact reference values at any spatial coordinate, thereby enabling precise, domain-wide quantification of structural deviations, including root mean square errors and maximum absolute local discrepancies. To systematically stress-test the modeling architectures, the evaluation framework is organized into three distinct physical configurations, uniformly referred to throughout this study as Case 1, Case 2, and Case 3. These configurations are intentionally engineered to emulate notoriously difficult fluid phenomena, ranging from strongly amplified near-wall boundary layers to sharp, non-trivial multiphase transition zones.

\subsection{Physical Boundary Conditions and Physics-Informed Optimization}

The ultimate fidelity of any physics-informed surrogate model depends fundamentally on accurately embedding hydrodynamic and thermodynamic constraints into the network optimization trajectory. For all investigated flow configurations, strict Dirichlet boundary conditions are extracted directly from the manufactured analytical solutions to enforce boundary consistency, yielding
\begin{equation}
\mathbf{U}(x,y)\big|_{\partial\Omega}=\mathbf{U}^{\ast}(x,y)\big|_{\partial\Omega},
\end{equation}
where $\partial\Omega$ defines the rigid boundaries of the computational domain. In addition to these boundary constraints, a sparse set of supervised interior collocation points is sampled from the exact reference fields to stabilize the underlying gradient descent process and establish a strictly fair comparative baseline across the distinct network architectures. Consequently, for any continuous neural approximation $\hat{\mathbf{U}}(x,y)$, the parameter optimization is guided by a composite, physics-constrained learning objective defined as
\begin{equation}
\mathcal{L}=\mathcal{L}_{\mathrm{data}}+\mathcal{L}_{\mathrm{bc}}+\mathcal{L}_{\mathrm{pde}}.
\end{equation}
Within this hybrid formulation, $\mathcal{L}_{\mathrm{data}}$ enforces structural adherence to the sampled interior fluid states, $\mathcal{L}_{\mathrm{bc}}$ aggressively penalizes deviations along the prescribed physical boundaries, and $\mathcal{L}_{\mathrm{pde}}$ minimizes the intrinsic physical residuals of the coupled Navier-Stokes, energy, and Poisson equations. This unified loss topology ensures that all evaluated models are subjected to identical physical conservation laws and data-driven supervision conditions during the training phase.

\section{Residual-Attention Physics-Informed Neural Network Architecture}
\label{sec:method}

\subsection{Baseline Physics-Informed Formulations and Network Backbones}

Physics-Informed Neural Networks (PINNs) approximate continuous fluid and thermodynamic variables by a neural network and enforce the governing equations through automatic differentiation \cite{raissi2019pinn,karniadakis2021piml}. For the present electro-thermoconvective problem, the input to the network is the spatial coordinate vector $\mathbf{x}=(x,y)$, and the output is the coupled field state
\begin{equation}
\hat{\mathbf{U}}(x,y)=\left[\hat{u}(x,y),\,\hat{v}(x,y),\,\hat{p}(x,y),\,\hat{T}(x,y),\,\hat{\phi}(x,y)\right].
\end{equation}

Given the predicted flow and thermal fields, spatial derivatives are rigorously obtained by automatic differentiation, and the corresponding partial differential equation residuals are expressed as
\begin{equation}
\mathbf{R}(x,y)=\left[R_{\mathrm{cont}},\,R_u,\,R_v,\,R_T,\,R_{\phi}\right].
\end{equation}
The standard PINN training objective is constructed by minimizing the discrepancy on supervised interior samples, the mismatch on physical boundary samples, and the residuals of the governing hydrodynamic equations. Therefore, the composite loss function takes the form
\begin{equation}
\mathcal{L}=\mathcal{L}_{\mathrm{data}}+\mathcal{L}_{\mathrm{bc}}+\mathcal{L}_{\mathrm{pde}},
\end{equation}
where the individual components are defined as
\begin{equation}
\mathcal{L}_{\mathrm{data}}=\frac{1}{N_d}\sum_{i=1}^{N_d}\left\|\hat{\mathbf{U}}(\mathbf{x}_i^d)-\mathbf{U}^{\ast}(\mathbf{x}_i^d)\right\|_2^2,
\end{equation}
\begin{equation}
\mathcal{L}_{\mathrm{bc}}=\frac{1}{N_b}\sum_{i=1}^{N_b}\left\|\hat{\mathbf{U}}(\mathbf{x}_i^b)-\mathbf{U}^{\ast}(\mathbf{x}_i^b)\right\|_2^2,
\end{equation}
and
\begin{equation}
\mathcal{L}_{\mathrm{pde}}=\frac{1}{N_r}\sum_{i=1}^{N_r}\left(
|R_{\mathrm{cont}}|^2+|R_u|^2+|R_v|^2+|R_T|^2+|R_{\phi}|^2
\right).
\end{equation}
Here, $N_d$, $N_b$, and $N_r$ denote the discrete numbers of interior supervised samples, boundary samples, and residual collocation points, respectively.

Although this formulation provides a mathematically straightforward learning objective, standard PINNs consistently struggle when the target flow fields contain strong near-wall amplification, annular abrupt multiphase transitions, or compact charged cores with sharp interfaces \cite{wang2021gradient,wang2022ntk,tancik2020fourier,sitzmann2020siren}. In such complex hydrodynamic situations, the network typically fits the smooth global background reasonably well but fails to reconstruct local critical regions with sufficient physical accuracy, suffering from severe numerical diffusion. This pervasive limitation directly motivates the introduction of enhanced network architectures capable of resolving localized gradients in electro-thermoconvective applications.

To provide a rigorous and fair quantitative evaluation, three distinct network architectures are benchmarked in this study, which are the pure PINN, the LSTM-PINN, and the proposed Residual-Attention PINN. All three models are strictly constrained by the same physical formulation, identical loss definitions, and the exact same training data configuration. Their fundamental difference lies exclusively in the representational capability of the network backbone.

The pure PINN adopts a standard fully connected multilayer perceptron to approximate the target field vector. This standard model serves as the foundational baseline architecture, representing the conventional physics-informed framework without any structural enhancement to capture sharp gradients.

The LSTM-PINN introduces a long short-term memory feature propagation mechanism into the continuous network backbone. Compared with the pure PINN, this recurrent model is deployed to potentially improve spatial feature interaction and hidden-state propagation. Within this study, the LSTM-PINN acts as a significantly stronger comparative baseline beyond the standard multilayer perceptron, a choice motivated by the original development of LSTM for sequential dependency modeling \cite{hochreiter1997lstm} and its recent exploratory applications in physics-informed fluid networks \cite{wang2025pirn}.

The proposed Residual-Attention PINN purposefully integrates both residual connections and attention enhancement into the feature propagation backbone. The residual structure is deliberately introduced to stabilize deep feature propagation and preserve coarse-scale, continuous background flow information \cite{he2016resnet}. Concurrently, the attention mechanism is designed to locally amplify informative patterns and strengthen the network response to spatially critical regions such as boundary layers, abrupt annular interfaces, and compact high-concentration charge structures \cite{vaswani2017attention}.

For all three evaluated models, the hidden feature dimension is fixed uniformly at 128 to ensure a matched parameter scale. The pure PINN uses a six-layer fully connected architecture with Tanh activation functions. The LSTM-PINN employs an input projection layer followed by two stacked single-layer LSTM modules and a final nonlinear output head. The proposed Residual-Attention PINN is constructed utilizing six stacked residual-attention blocks maintaining the identical hidden width, which is ultimately connected to a nonlinear output head.

\subsection{Residual-Attention Mechanism for Localized Gradient Resolution}

\begin{figure*}[t]
    \centering
    \includegraphics[width=0.96\textwidth]{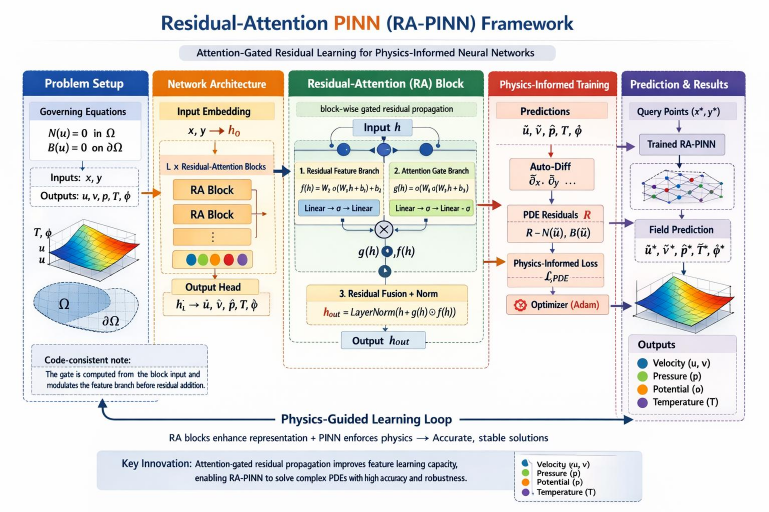}
    \caption{Schematic of the proposed Residual-Attention Physics-Informed Neural Network (RA-PINN) for multiscale flow reconstruction[cite: 595, 669]. The framework maps spatial coordinates $(x,y)$ to the coupled electro-thermoconvective state variables $(u,v,p,T,\phi)$, enforcing the governing Navier-Stokes and Poisson equations via an automatic differentiation engine[cite: 596, 597, 670, 671]. Each computational block integrates a residual-feature branch to stabilize global background propagation and a parallel attention-gate branch to adaptively sensitize the network to steep physical gradients[cite: 599, 600, 672]. This dual-pathway architecture is specifically engineered to resolve localized extreme structures, such as viscous boundary layers and sharp interfacial charge topologies, without triggering the numerical diffusion typically inherent in standard physics-informed models[cite: 598, 601, 674].}
    \label{fig:ra_pinn_architecture}
\end{figure*}

Figure \ref{fig:ra_pinn_architecture} illustrates the comprehensive architecture of the proposed Residual-Attention PINN. The predictive model takes the spatial coordinates $(x,y)$ as a continuous input and predicts the five coupled physical fields, specifically the velocity components $u$ and $v$, pressure $p$, temperature $T$, and electric potential $\phi$. These predicted physical fields are subsequently differentiated through an automatic differentiation engine to construct the exact partial differential equation residuals required by the physics-informed optimization loop. Within the core backbone network, multiple residual-attention blocks are sequentially stacked to drastically improve the high-fidelity representation of difficult local structures. As explicitly detailed in the residual-attention module of Fig.\ref{fig:ra_pinn_architecture}, each computational block contains a dedicated residual-feature branch and a parallel attention-gate branch. The modulation gate is dynamically generated from the block input and is utilized to adaptively weight the feature branch before executing the residual addition, which is then immediately followed by robust layer normalization. Through this highly specialized design, the proposed network successfully preserves stable background field information while simultaneously enhancing localized sensitivity to strongly amplified fluid structures.

Let $\mathbf{h}^{(l)}$ mathematically denote the hidden feature state at layer $l$. Within the residual-attention backbone, the nonlinear feature branch of a specific block is written as
\begin{equation}
\mathbf{f}^{(l)}=\mathcal{F}\left(\mathbf{h}^{(l)};\Theta_f^{(l)}\right),
\end{equation}
where $\mathcal{F}(\cdot)$ represents the residual-feature transformation parameterized by the weight set $\Theta_f^{(l)}$. In the present computational implementation, this primary branch is realized by two fully connected linear transformations coupled with nonlinear activation functions. This residual learning paradigm effectively alleviates optimization degradation and facilitates smooth continuous information propagation across deep layers \cite{he2016resnet}.

To specifically address and enhance local feature sensitivity, the attention-gate branch operates entirely in parallel. The spatial gate is generated directly from the incoming block input and is expressed algebraically as
\begin{equation}
\mathbf{g}^{(l)}=\sigma\!\left(\mathcal{G}\left(\mathbf{h}^{(l)};\Theta_g^{(l)}\right)\right),
\end{equation}
where $\mathcal{G}(\cdot)$ denotes the gate-generating transformation, $\Theta_g^{(l)}$ is the corresponding learnable parameter set, and $\sigma(\cdot)$ denotes the sigmoid activation function used to produce strictly bounded, channel-wise modulation coefficients. The subsequent gated residual interaction is then formulated as
\begin{equation}
\mathbf{h}^{(l+1)}=\mathrm{LayerNorm}\!\left(\mathbf{h}^{(l)}+\mathbf{g}^{(l)}\odot \mathbf{f}^{(l)}\right),
\end{equation}
where $\odot$ defines the element-wise multiplication operator. In this specific mathematical form, the attention gate avoids directly reweighting the main hidden feature itself; instead, it adaptively modulates the extracted residual-feature branch immediately prior to the residual addition. This precise mathematical mechanism forces the network to place significantly more optimization emphasis on regions or geometric patterns that are inherently more difficult to model, such as near-wall viscous boundary layers, abrupt circular transitions, and highly concentrated charge core structures \cite{vaswani2017attention}.

In the complete proposed architecture, the Cartesian input coordinates are initially projected into the expanded hidden feature space, seamlessly propagated through six stacked residual-attention blocks, and ultimately mapped to the five target fluid variables through the final output head. Therefore, the proposed RA-PINN architecture can be physically interpreted as a gated residual feature propagation network perfectly embedded into the standard physics-informed continuous learning framework. Compared with a plain multilayer perceptron, this dual-branch design is uniquely intended to preserve global structural consistency across the entire flow domain while radically improving the localized representation of difficult, steep-gradient regions.

This deliberate structural combination is particularly suitable for the present electrohydrodynamic benchmark set, primarily because all selected configurations contain strongly localized amplification, annular abrupt transitions, or compact high-concentration cores. Specifically, Case 1 simulates an exponential electric-charge boundary layer directly near the electrode at $x=1$, which stringently requires the model to capture a strongly amplified near-wall momentum response alongside transverse modulation. Case 2 encompasses a ring-shaped abrupt interface induced by a central cylindrical electrode, which rigorously tests whether the neural network can successfully preserve radially localized annular transitions without introducing artificial smoothing. Finally, Case 3 features a circular sharp-interface field enclosing a strong local charge concentration, which inherently challenges the predictive model to reconstruct a dense, compact charged core together with sharp radial topological variations.

Consequently, for these highly localized hydrodynamic problems, a single plain multilayer perceptron inherently tends to erroneously smooth out local extreme fluid features during optimization due to pervasive spectral bias. Standard residual connections fundamentally help maintain the stable propagation of continuous feature information across layers, which is highly useful for preserving the broader global flow structure. However, global stability alone remains completely insufficient for capturing highly localized physical patterns. The integrated gated attention branch decisively compensates for this specific limitation by dynamically increasing the mathematical sensitivity of the network to difficult local flow regions through adaptive feature modulation. Therefore, the sophisticated combination of residual learning and gated attention weighting is expected to be substantially more effective than utilizing a plain multilayer perceptron when the target fluid solution contains viscous boundary layers, annular abrupt interfaces, or localized compact sharp-interface charge structures.

\subsection{Physics-Constrained Optimization and Evaluation Strategy}

All selected models are strictly trained under the exact same optimization framework in order to ensure a fully unbiased and fair quantitative comparison. The comprehensive training samples consist of three distinct parts, which are interior supervised points generated directly from the analytical reference solution, explicit boundary points extracted from the exact Dirichlet boundary conditions, and interior residual collocation points continuously used to strictly enforce the coupled governing partial differential equations. The unified total loss is thereby defined by
\begin{equation}
\mathcal{L}_{\mathrm{total}}=
\mathcal{L}_{\mathrm{data}}+
\mathcal{L}_{\mathrm{bc}}+
\mathcal{L}_{\mathrm{pde}}.
\end{equation}

In computational practice, the continuous PDE residual loss can also be systematically written as the direct sum of individual physical residual components, expressed as
\begin{equation}
\mathcal{L}_{\mathrm{pde}}=
\mathcal{L}_{\mathrm{cont}}+
\mathcal{L}_{u}+
\mathcal{L}_{v}+
\mathcal{L}_{T}+
\mathcal{L}_{\phi},
\end{equation}
where the five distinct terms correspondingly map to the fluid continuity, two-dimensional momentum, thermal energy, and electrostatic potential equations. This explicit physical decomposition is extraordinarily useful for continuously monitoring the dynamic training process and accurately identifying whether a given model architecture demonstrates optimization difficulty within a specific physical component of the coupled fluid system.

The diverse network parameters are optimized utilizing the exact same numerical optimizer and an identical training learning-rate schedule across all three tested architectures. Consequently, the resulting quantitative comparison focuses exclusively on the independent physical effect of the underlying network backbone rather than completely arbitrary differences in optimization training hyperparameters. The final predictive evaluation is rigorously conducted by comparing the neural-predicted flow fields with the exact analytical reference solutions primarily in terms of globally averaged error, localized maximum absolute error, overall structural similarity, and iterative convergence behavior.

The overall computational workflow proceeds systematically to establish the predictive methodology. First, the benchmark-specific analytical reference solution is mathematically constructed, and the corresponding continuous source terms are explicitly derived. Subsequently, interior supervised discrete samples, physical boundary boundary points, and dense residual collocation points are uniformly generated across the spatial domain. The raw spatial coordinates are then fed dynamically into the neural network to actively predict the coupled fluid and thermodynamic variables. Following this forward prediction, spatial derivatives are precisely computed via the automatic differentiation engine to accurately evaluate the continuous partial differential equation residuals. The underlying network parameters are then iteratively optimized by minimizing the combined data, boundary, and physical residual losses via gradient descent. Finally, the predicted flow fields from the different competing architectures are quantitatively compared against the exact manufactured reference fields to definitively assess structural fidelity and local gradient resolution.

This highly structured workflow is completely shared by the pure PINN baseline, the recurrent LSTM-PINN, and the proposed Residual-Attention PINN. Therefore, any observable difference in the final multiphysics prediction performance can be directly and exclusively attributed to the fundamental representational capability of the neural network architecture itself. In the subsequent section, the three distinct electro-thermoconvective benchmark cases and their corresponding explicit experimental settings are introduced and analyzed in extensive detail.

\section{Electrohydrodynamic Benchmark Configurations and Computational Setup}
\label{sec:benchmark}

\subsection{Physical Configurations for Gradient and Interface Resolution}

To keep the specific naming convention consistent throughout the comparative analysis, the analytical benchmark set is organized directly as Case 1, Case 2, and Case 3. These three distinct flow configurations perfectly correspond to three complementary forms of local structural difficulty frequently encountered in computational fluid dynamics. Specifically, the test matrix encompasses a near-electrode exponential boundary layer characterized by strong transverse modulation, a radially localized annular abrupt transition zone, and a highly concentrated charged core enclosed by a sharp interfacial field. Together, these carefully constructed physical scenarios provide a highly focused benchmark set designed for testing the intrinsic ability of different neural architectures to recover extreme local high-contrast fluid structures under a unified coupled-physics formulation.

\subsection{Case 1: Exponential Electro-Convective Boundary Layer}

The first configuration, designated as Case 1, is meticulously designed to emulate an exponential electric-charge boundary layer dynamically evolving near a solid electrode located at $x=1$. The prescribed analytical flow field incorporates exponentially amplified terms of the form $\exp[\beta(x-1)]$ subjected to transverse modulation along the $y$-axis, forcing the most intense spatial variations to concentrate strictly along the right physical boundary. This specific physical scenario presents a significant computational hurdle because the multi-physics field remains globally smooth over the majority of the fluid domain while simultaneously exhibiting explosive near-wall gradient amplification within an extremely narrow spatial band. A standard neural surrogate model optimized primarily for global averaged metrics typically diffuses this critical local boundary-layer structure and severely underestimates its physical intensity. Therefore, Case 1 rigorously tests whether the network architecture can preserve the sheer sharpness and strict spatial localization of a strongly amplified electro-convective near-wall response.

\subsection{Case 2: Annular Abrupt Interface Induced by a Cylindrical Electrode}

Building upon the boundary-layer challenges, Case 2 introduces a fundamentally different hydrodynamic topological challenge by modeling a localized annular abrupt interface generated around a central cylindrical electrode. The corresponding analytical benchmark field is constructed utilizing coupled radial variables and continuously differentiable annular-shell functions, ensuring the target thermodynamic and fluid fields exhibit a distinct ring-shaped transition rather than a simplified planar or axis-aligned discontinuity. The intrinsic difficulty of this configuration lies entirely in the accurate spatial reconstruction of a highly localized annular transition zone possessing a nontrivial radial structure. If the underlying network architecture lacks adequate representational power to capture steep radial gradients, the simulated multiphase interface will become overly diffused, its physical radius will distort, and the localized transition intensity will categorically degrade. Consequently, Case 2 serves as an optimal testbed for evaluating the capacity of the proposed architecture to preserve strictly non-planar interface geometries within radially constrained electrohydrodynamic environments.

\subsection{Case 3: Compact Charged Core with Sharp Interfacial Topology}

The final configuration, denoted as Case 3, features a highly compact circular charged-core structure governed by a sharp radial switching function. The exact physical solution imposes a tightly localized charge distribution and a strictly circular sharp-interface topology seamlessly embedded into the surrounding coupled velocity, pressure, temperature, and electric-potential fields. Compared with the prior boundary layer and annular transition scenarios, Case 3 critically emphasizes the complex physical interaction between a highly dense local concentration and a distinct circular multiphase boundary. The primary numerical challenge here is to mathematically preserve the strict spatial integrity of the central charged region while simultaneously maintaining the continuous physical behavior of the surrounding broader flow field. This specific flow scenario therefore precisely evaluates the capability of the network to model highly compact, strongly localized electro-thermal structures without triggering spurious numerical oscillations or suffering from excessive domain-wide smoothing.

\subsection{Computational Setup and Training Parameters}

All three electrohydrodynamic configurations are computationally resolved on an identical square physical domain defined spatially by $\Omega=[-1,1]\times[-1,1]$. To drive the physics-informed optimization, the supervised interior state observations are systematically generated on a dense structured spatial grid, with the total dataset strictly partitioned into training and validation subsets utilizing a ratio of $7{:}3$. Physical boundary constraints are directly extracted from the analytical reference solutions, while the partial differential equation residuals are dynamically evaluated at uniformly distributed interior collocation points. To guarantee a strictly unbiased comparative analysis, the pure PINN, LSTM-PINN, and RA-PINN models are fundamentally constrained to utilize the exact same hidden layer width, numerical optimizer, computational training budget, and validation protocol. 

The comprehensive hyperparameter settings dictating this rigorous experimental framework are explicitly summarized in Table~\ref{tab:exp_settings}.

\begin{table}[htbp]
\caption{Summary of numerical configurations and shared hyperparameters for the physics-informed training across all benchmark scenarios.}
\label{tab:exp_settings}
\centering
\begin{ruledtabular}
\begin{tabular}{ll}
\textbf{Numerical Configuration Item} & \textbf{Specification} \\
\hline
Computational domain & $[-1,1] \times [-1,1]$  \\
Training/validation split ratio & $7:3$  \\
Hidden layer feature dimension & $128$ (for all three backbones)  \\
Optimizer and total iterations & Adam, $50,000$ steps  \\
Interior collocation points ($N_r$) & $16,000$  \\
Dirichlet boundary samples ($N_b$) & $1,436$  \\
\end{tabular}
\end{ruledtabular}
\end{table}

\section{Results and Physical Discussion}
\label{sec:results}

\subsection{Overall Evaluation of Field Reconstruction Fidelity}

The predictive capabilities of the three tested models are rigorously evaluated based on their averaged domain accuracy, localized maximum error, qualitative physical field reconstruction, and iterative convergence behavior. Across all three electrohydrodynamic configurations, the proposed RA-PINN framework achieves the most robust quantitative performance. The integrated residual pathway inherently stabilizes continuous flow feature propagation, while the parallel attention-gate branch dramatically improves numerical sensitivity to steep localized gradients and sharp multiphase transitions. 

At the aggregate level, the statistical analysis reveals a consistent performance hierarchy of RA-PINN exceeding LSTM-PINN, which in turn exceeds the pure PINN baseline across all tested scenarios. In the first configuration, the domain-averaged RMSE of the RA-PINN is drastically reduced from $3.538\times10^{-4}$ to $7.559\times10^{-5}$ relative to the pure PINN, and from $1.910\times10^{-4}$ to $7.559\times10^{-5}$ relative to the recurrent LSTM-PINN. In the second configuration, the error reduction is even more pronounced, with the averaged RMSE plummeting from $5.525\times10^{-3}$ and $1.892\times10^{-3}$ down to $4.217\times10^{-4}$. For the third configuration, the RA-PINN again yields the lowest statistical error while delivering the best overall preservation of the compact charged-core topology. These comprehensive results confirm that the residual-attention architecture successfully eliminates severe numerical diffusion, improving both global thermodynamic consistency and local hydrodynamic structural fidelity.

\begin{table}[htbp]
\caption{Overall quantitative performance comparison of pure PINN, LSTM-PINN, and RA-PINN backbones across the three canonical configurations. Error metrics represent the arithmetic mean across the five coupled electro-thermoconvective fields.}
\label{tab:overall_results}
\centering
\begin{ruledtabular}
\begin{tabular}{llccc}
\textbf{Scenario} & \textbf{Model Backbone} & \textbf{Avg. RMSE} & \textbf{Avg. MAE} & \textbf{Avg. $L_{\infty}$} \\
\hline
Case 1 & pure PINN & 3.538e-04 & 2.728e-04 & 2.430e-03 \\
& LSTM-PINN & 1.910e-04 & 1.356e-04 & 1.783e-03 \\
& RA-PINN & 7.559e-05 & 5.770e-05 & 5.059e-04 \\
\hline
Case 2 & pure PINN & 5.525e-03 & 4.279e-03 & 2.377e-02 \\
& LSTM-PINN & 1.892e-03 & 1.355e-03 & 1.113e-02 \\
& RA-PINN & 4.217e-04 & 3.244e-04 & 2.745e-03 \\
\hline
Case 3 & pure PINN & 5.909e-03 & 4.372e-03 & 2.716e-02 \\
& LSTM-PINN & 2.914e-03 & 2.137e-03 & 1.508e-02 \\
& RA-PINN & 1.299e-03 & 8.552e-04 & 1.481e-02 \\
\end{tabular}
\end{ruledtabular}
\end{table}

\subsection{Resolution of the Electro-Convective Boundary Layer (Case 1)}

For the first physical configuration, the central computational challenge is the accurate recovery of the highly amplified exponential boundary layer situated near the solid electrode at $x=1$. In this specific hydrodynamic situation, a reliable surrogate model must correctly represent both the broad convective background and the extremely narrow near-wall region where the electrostatic response is maximally amplified. The pure PINN baseline successfully reproduces the smooth global field but severely blurs the near-wall gradient, effectively introducing an artificial thickening of the boundary layer that physically equates to an underprediction of near-wall electrohydrodynamic forcing. The LSTM-PINN slightly alleviates this diffusion issue, yet an unacceptable local mismatch near the most amplified spatial zone still persists.

The quantitative error metrics explicitly confirm the architectural advantage of the proposed method. The RA-PINN attains the lowest error boundary for every coupled field in Case 1, reducing the averaged RMSE by approximately $78.6\%$ relative to the pure PINN and by $60.4\%$ relative to the LSTM-PINN. This statistical improvement is overwhelmingly concentrated in the near-electrode high-gradient region, where preserving the sheer sharpness of the boundary layer is physically essential for calculating accurate wall shear stresses and charge distributions. This specific scenario therefore provides direct, unambiguous evidence that residual-attention enhancement is highly effective for capturing electrode-adjacent exponential flow structures.

\begin{figure*}[t]
    \centering
    \includegraphics[width=\textwidth]{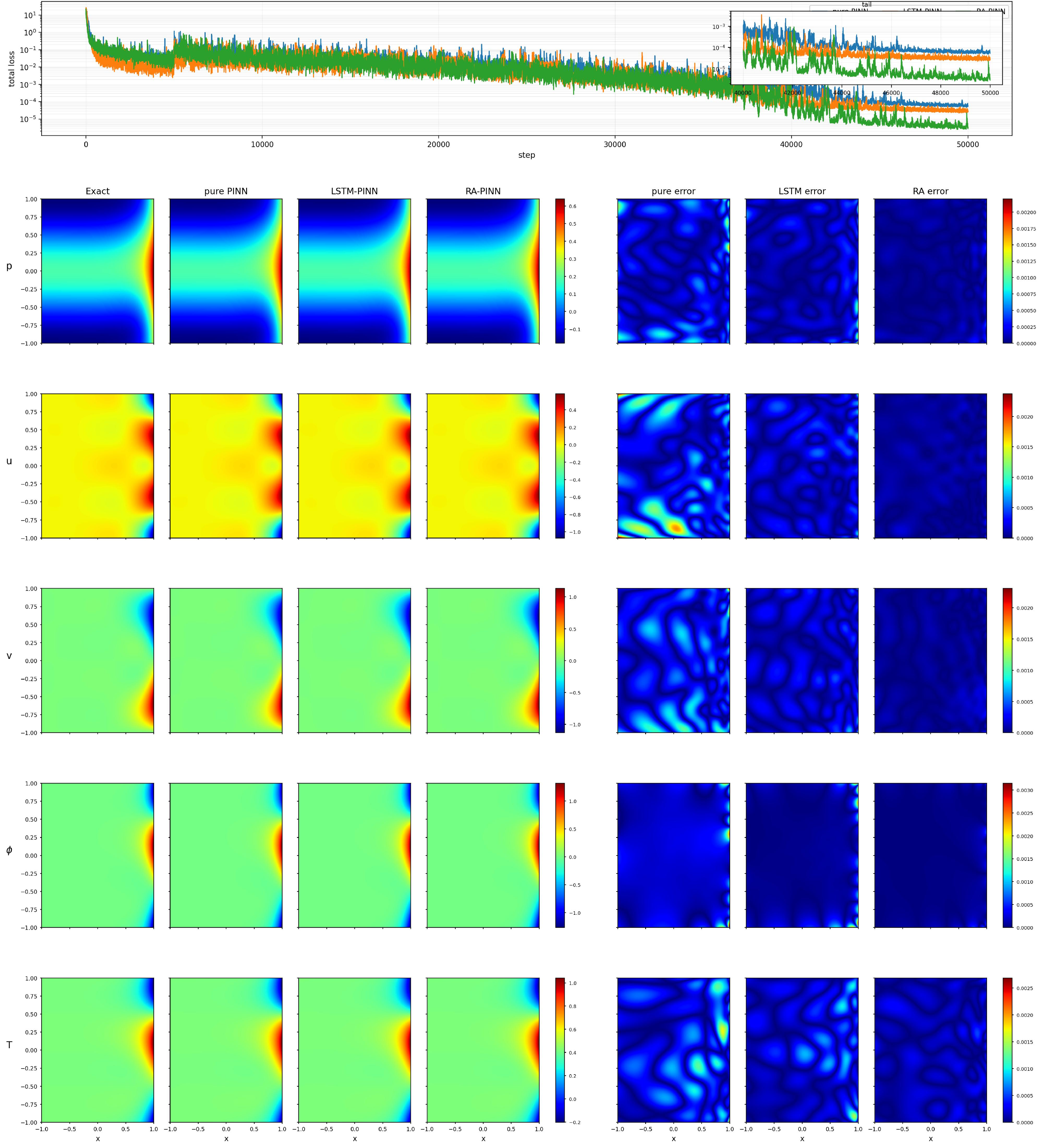}
    \caption{Comparative reconstruction of the exponential electro-convective boundary layer (Case 1) under the unified physics-informed benchmark. The top row displays the total loss convergence curves across the 50,000-step training trajectory for the three backbones. The subsequent columns present the exact reference fields, the neural predictions from pure PINN, LSTM-PINN, and RA-PINN, and the corresponding absolute error distributions for the coupled pressure $p$, velocity components $(u, v)$, electric potential $\phi$, and temperature $T$. Note that the RA-PINN successfully resolves the steep gradients near the electrode at $x=1$ with minimal error concentration, whereas the baselines exhibit significant numerical diffusion and distortion in the boundary-layer zone.}
    \label{fig:case1_visual}
\end{figure*}

\begin{table*}[htbp]
\caption{Detailed field-wise error metrics for the exponential electro-convective boundary layer (Case 1). The comparison highlights the superior resolution of localized gradients achieved by the RA-PINN architecture across all coupled physical variables.}
\label{tab:case1_metrics}
\centering
\begin{ruledtabular}
\begin{tabular}{lcccccc}
\textbf{Model Backbone} & \textbf{Field} & \textbf{RMSE} & \textbf{MSE} & \textbf{MAE} & \textbf{$L_2$ error} & \textbf{$L_{\infty}$ error} \\
\hline
pure PINN & $u$ & 4.704e-04 & 2.213e-07 & 3.505e-04 & 3.740e-03 & 2.375e-03  \\
& $v$ & 3.522e-04 & 1.240e-07 & 2.811e-04 & 2.087e-03 & 2.308e-03  \\
& $p$ & 2.517e-04 & 6.334e-08 & 1.854e-04 & 1.568e-03 & 1.635e-03  \\
& $T$ & 4.067e-04 & 1.654e-07 & 3.099e-04 & 9.459e-04 & 2.685e-03  \\
& $\phi$ & 2.881e-04 & 8.299e-08 & 2.371e-04 & 1.828e-03 & 3.146e-03  \\
\hline
LSTM-PINN & $u$ & 1.838e-04 & 3.379e-08 & 1.458e-04 & 1.462e-03 & 1.444e-03  \\
& $v$ & 1.610e-04 & 2.591e-08 & 1.305e-04 & 9.538e-04 & 1.080e-03  \\
& $p$ & 1.481e-04 & 2.193e-08 & 1.057e-04 & 9.227e-04 & 2.198e-03  \\
& $T$ & 2.776e-04 & 7.705e-08 & 2.040e-04 & 6.456e-04 & 1.509e-03  \\
& $\phi$ & 1.845e-04 & 3.402e-08 & 9.209e-05 & 1.171e-03 & 2.682e-03  \\
\hline
RA-PINN & $u$ & 6.748e-05 & 4.553e-09 & 5.772e-05 & 5.365e-04 & 4.735e-04  \\
& $v$ & 6.078e-05 & 3.694e-09 & 4.714e-05 & 3.602e-04 & 3.388e-04  \\
& $p$ & 5.918e-05 & 3.502e-09 & 4.355e-05 & 3.687e-04 & 4.148e-04  \\
& $T$ & 1.330e-04 & 1.769e-08 & 1.089e-04 & 3.093e-04 & 5.180e-04  \\
& $\phi$ & 5.752e-05 & 3.308e-09 & 3.120e-05 & 3.651e-04 & 7.842e-04  \\
\end{tabular}
\end{ruledtabular}
\end{table*}

\subsection{Preservation of the Annular Multiphase Interface (Case 2)}

The primary difficulty embedded within the second flow configuration is the strict topological reconstruction of a ring-shaped abrupt transition zone. Unlike a simplified planar or single-direction fluid discontinuity, the annular interface demands that the neural model simultaneously preserve the intrinsic radial geometry, the exact physical transition width, and the localized interfacial intensity. Standard purely connected PINNs generally capture the broad radial trend but inevitably produce an over-smoothed interfacial ring, effectively introducing severe artificial numerical diffusion that distorts the effective transition radius.

The computed spatial metrics precisely verify this deleterious smoothing behavior in the baselines. In stark contrast, the RA-PINN architecture yields the lowest spatial error for all five coupled fields in Case 2, successfully reducing the averaged RMSE by approximately $92.4\%$ relative to the pure PINN and by $77.7\%$ relative to the LSTM-PINN. This massive performance gain is most highly pronounced within this specific annular-interface benchmark, strongly indicating that the integrated residual-attention mechanism is uniquely equipped to preserve complex radial interface geometries while decisively suppressing localized error accumulation along steep transition boundaries.

\begin{figure*}[t]
    \centering
    \includegraphics[width=\textwidth]{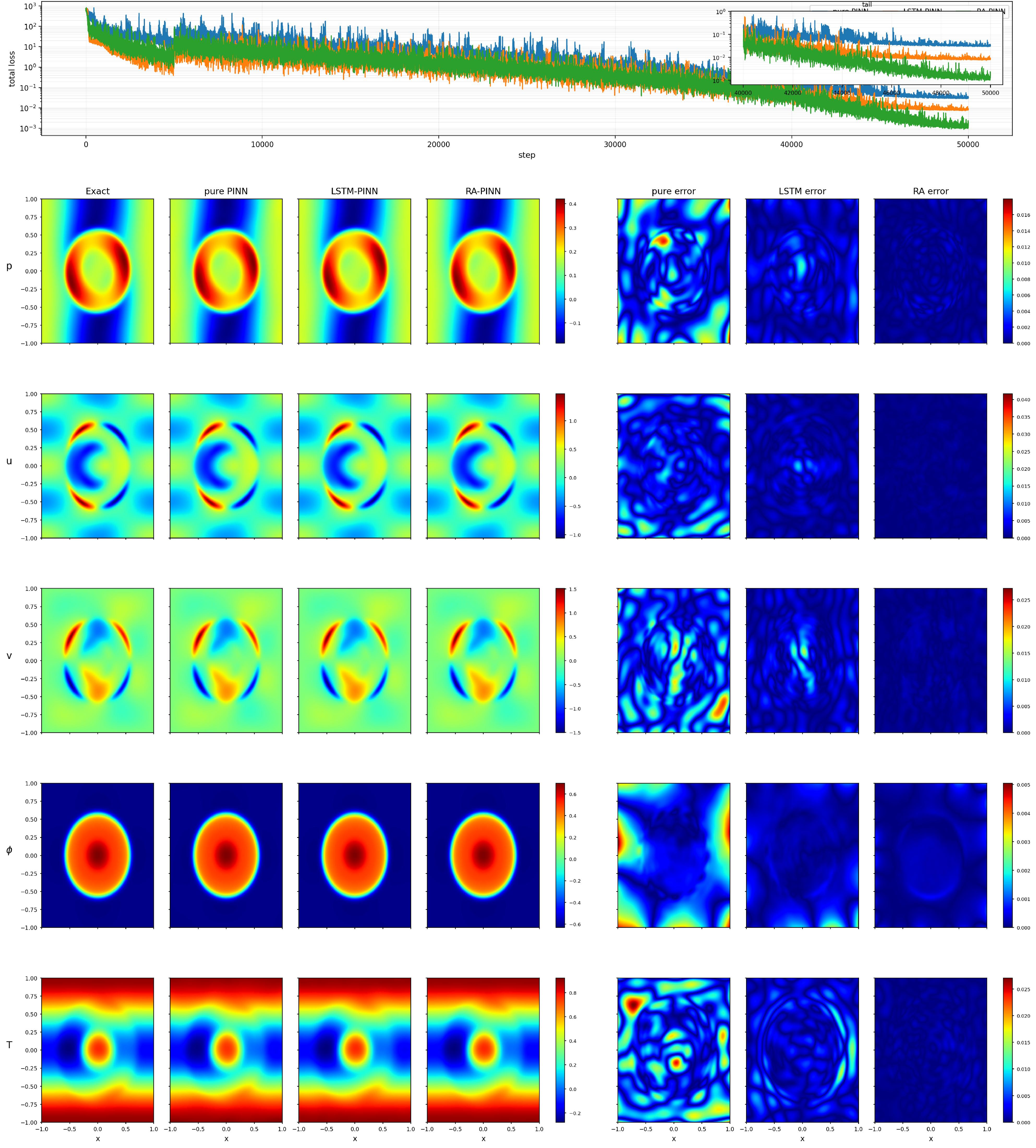}
    \caption{Numerical reconstruction of the annular abrupt interface induced by a cylindrical electrode (Case 2). The top panel illustrates the iterative reduction of the composite physics-informed loss for the evaluated architectures. The field-wise matrix compares the analytical reference solutions (first column) with the predictions and absolute error distributions for pressure $p$, velocity components $(u, v)$, electric potential $\phi$, and temperature $T$. The results highlight the capacity of the RA-PINN to maintain the sharp radial transition and interfacial curvature without the excessive numerical diffusion observed in the standard MLP and recurrent baselines.}
    \label{fig:case2_visual}
\end{figure*}

\begin{table*}[htbp]
\caption{Detailed field-wise error metrics for the annular abrupt interface configuration (Case 2). The quantitative results underscore the significant reduction in interfacial numerical diffusion provided by the RA-PINN across all electro-thermoconvective variables.}
\label{tab:case2_metrics}
\centering
\begin{ruledtabular}
\begin{tabular}{lcccccc}
\textbf{Model Backbone} & \textbf{Field} & \textbf{RMSE} & \textbf{MSE} & \textbf{MAE} & \textbf{$L_2$ error} & \textbf{$L_{\infty}$ error} \\
\hline
pure PINN & $u$ & 7.372e-03 & 5.434e-05 & 5.603e-03 & 2.398e-02 & 4.165e-02 \\
& $v$ & 6.375e-03 & 4.064e-05 & 4.988e-03 & 2.225e-02 & 2.713e-02 \\
& $p$ & 4.588e-03 & 2.105e-05 & 3.488e-03 & 2.623e-02 & 1.796e-02 \\
& $T$ & 7.845e-03 & 6.155e-05 & 6.226e-03 & 1.424e-02 & 2.705e-02 \\
& $\phi$ & 1.444e-03 & 2.085e-06 & 1.090e-03 & 2.534e-03 & 5.045e-03 \\
\hline
LSTM-PINN & $u$ & 1.941e-03 & 3.769e-06 & 1.389e-03 & 6.316e-03 & 1.418e-02 \\
& $v$ & 2.384e-03 & 5.683e-06 & 1.633e-03 & 8.319e-03 & 1.552e-02 \\
& $p$ & 1.499e-03 & 2.247e-06 & 1.126e-03 & 8.571e-03 & 7.529e-03 \\
& $T$ & 3.183e-03 & 1.013e-05 & 2.309e-03 & 5.776e-03 & 1.562e-02 \\
& $\phi$ & 4.517e-04 & 2.041e-07 & 3.162e-04 & 7.927e-04 & 2.777e-03 \\
\hline
RA-PINN & $u$ & 4.321e-04 & 1.868e-07 & 3.311e-04 & 1.406e-03 & 4.224e-03 \\
& $v$ & 5.096e-04 & 2.597e-07 & 3.997e-04 & 1.779e-03 & 3.646e-03 \\
& $p$ & 4.198e-04 & 1.762e-07 & 3.116e-04 & 2.400e-03 & 2.587e-03 \\
& $T$ & 5.346e-04 & 2.858e-07 & 4.204e-04 & 9.703e-04 & 2.177e-03 \\
& $\phi$ & 2.124e-04 & 4.511e-08 & 1.591e-04 & 3.727e-04 & 1.092e-03 \\
\end{tabular}
\end{ruledtabular}
\end{table*}

\subsection{Reconstruction of the Compact Charged Core (Case 3)}

The final hydrodynamic configuration introduces the complex challenge of simultaneously resolving a highly compact circular charged region actively interacting with a sharp interfacial boundary. In this computationally demanding scenario, the predictive model must flawlessly preserve the spatial integrity of the dense local charged core without disrupting the broader continuity of the surrounding physical domain. This specific modeling requirement is exceedingly rigorous because both the extreme compactness of the charge and the surrounding multiphase interface sharpness must be strictly maintained to prevent the onset of unphysical flow oscillations.

The quantitative physical comparison unequivocally supports the efficacy of the proposed architecture in meeting this strict demand. The RA-PINN definitively achieves the absolute best average RMSE and MAE throughout Case 3, showcasing a remarkable averaged RMSE reduction of approximately $78.0\%$ relative to the pure PINN baseline and $55.4\%$ relative to the recurrent LSTM-PINN. Although the compact charged-core structure mathematically remains a more formidable challenge than the simple boundary layer in Case 1, the residual-attention design proves highly resilient, providing the most physically reliable preservation of the localized charged region and its surrounding sharp interface.

\begin{figure*}[t]
    \centering
    \includegraphics[width=\textwidth]{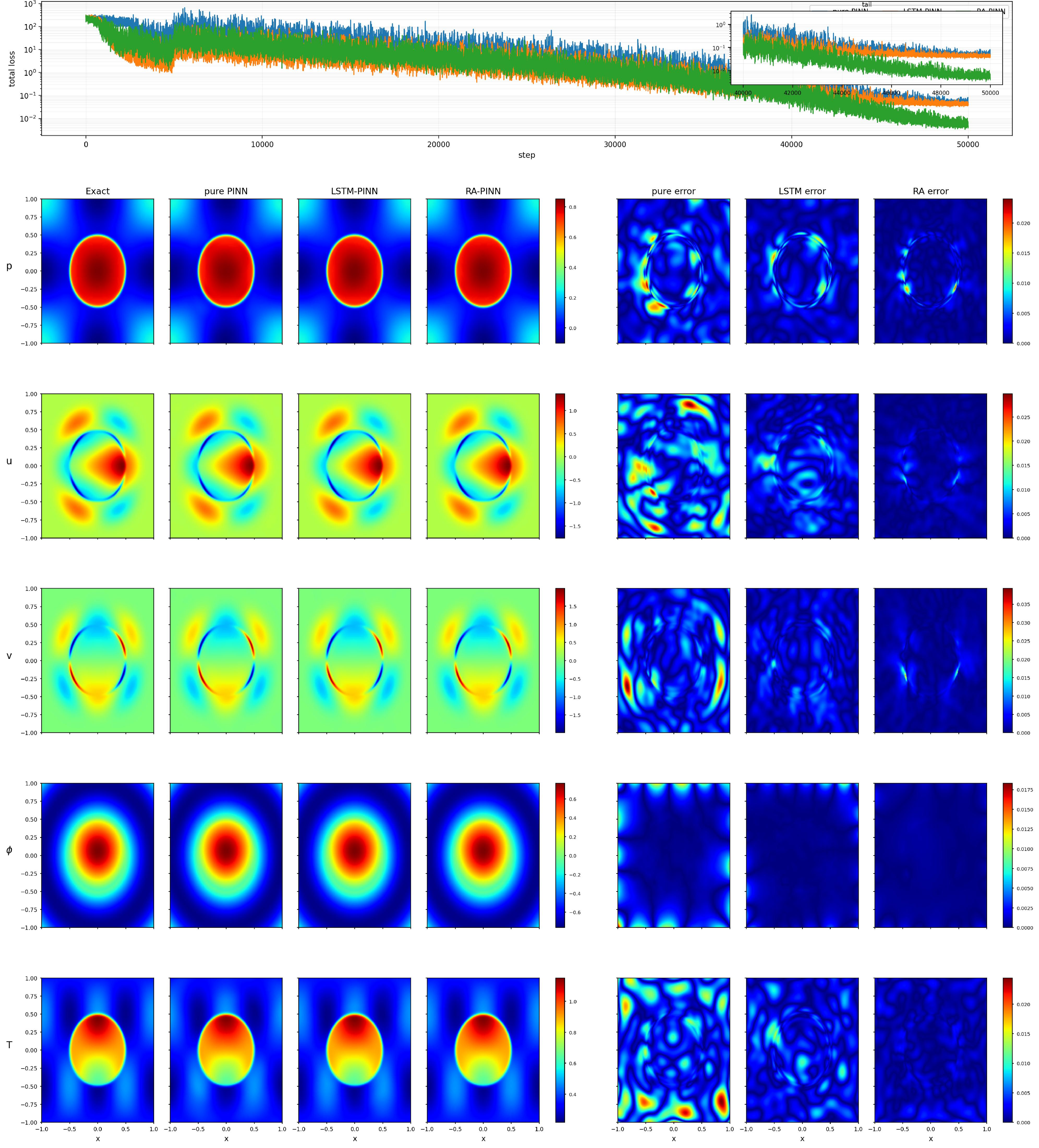}
    \caption{Numerical reconstruction of the compact charged core with sharp interfacial topology (Case 3). The top panel displays the convergence behavior of the physics-informed loss components during the optimization process. The subsequent field-wise comparison illustrates the exact reference solutions, the neural predictions from the three backbones, and the absolute error distributions for pressure $p$, velocity $(u, v)$, electric potential $\phi$, and temperature $T$. The results demonstrate that the RA-PINN successfully preserves the spatial integrity of the high-concentration charged core and the surrounding sharp radial transitions, whereas the baselines exhibit substantial smoothing and unphysical artifacts in the localized extreme zones.}
    \label{fig:case3_visual}
\end{figure*}

\begin{table*}[htbp]
\caption{Detailed field-wise error metrics for the compact charged core with sharp interfacial topology (Case 3). The comparative data highlights the robustness of the RA-PINN in accurately resolving highly localized physical features and steep gradients within the coupled flow field.}
\label{tab:case3_metrics}
\centering
\begin{ruledtabular}
\begin{tabular}{lcccccc}
\textbf{Model Backbone} & \textbf{Field} & \textbf{RMSE} & \textbf{MSE} & \textbf{MAE} & \textbf{$L_2$ error} & \textbf{$L_{\infty}$ error} \\
\hline
pure PINN & $u$ & 8.037e-03 & 6.459e-05 & 6.231e-03 & 2.163e-02 & 2.980e-02 \\
& $v$ & 8.058e-03 & 6.494e-05 & 5.837e-03 & 2.374e-02 & 3.928e-02 \\
& $p$ & 4.373e-03 & 1.912e-05 & 3.132e-03 & 1.258e-02 & 2.400e-02 \\
& $T$ & 6.615e-03 & 4.375e-05 & 5.061e-03 & 1.299e-02 & 2.443e-02 \\
& $\phi$ & 2.464e-03 & 6.073e-06 & 1.597e-03 & 4.355e-03 & 1.831e-02 \\
\hline
LSTM-PINN & $u$ & 3.963e-03 & 1.570e-05 & 2.973e-03 & 1.066e-02 & 1.696e-02 \\
& $v$ & 3.546e-03 & 1.257e-05 & 2.716e-03 & 1.045e-02 & 1.670e-02 \\
& $p$ & 2.767e-03 & 7.654e-06 & 1.877e-03 & 7.957e-03 & 1.813e-02 \\
& $T$ & 3.296e-03 & 1.086e-05 & 2.564e-03 & 6.473e-03 & 1.390e-02 \\
& $\phi$ & 1.000e-03 & 1.001e-06 & 5.549e-04 & 1.768e-03 & 9.720e-03 \\
\hline
RA-PINN & $u$ & 1.364e-03 & 1.861e-06 & 9.118e-04 & 3.671e-03 & 1.433e-02 \\
& $v$ & 1.888e-03 & 3.566e-06 & 1.219e-03 & 5.565e-03 & 3.106e-02 \\
& $p$ & 1.567e-03 & 2.456e-06 & 8.324e-04 & 4.507e-03 & 2.117e-02 \\
& $T$ & 1.286e-03 & 1.654e-06 & 1.014e-03 & 2.526e-03 & 5.131e-03 \\
& $\phi$ & 3.903e-04 & 1.524e-07 & 2.995e-04 & 6.898e-04 & 2.341e-03 \\
\end{tabular}
\end{ruledtabular}
\end{table*}

\subsection{Summary of Physical Reconstruction Capabilities}

Taken together, the comprehensive quantitative and qualitative analyses categorically confirm that the architectural advantage of the RA-PINN is not artificially restricted to a single fluid topology. Instead, this enhanced predictive capability consistently extends across intensely amplified boundary-layer structures, annular abrupt multiphase interfaces, and tightly localized circular sharp-interface fields. This unyielding physical consistency strongly establishes that the integrated residual-attention mechanism provides a highly robust computational methodology for eliminating artificial numerical diffusion and maximizing local structural fidelity in complex electro-thermoconvective modeling scenarios.

\section{Physical Implications for Electro-Thermoconvective Flows}
\label{sec:engineering}

The physical implications of this study extend significantly beyond the immediate mathematical validation of a neural architecture. In complex electro-thermoconvective flows, the overarching macroscopic physical behavior is fundamentally governed by highly localized spatial structures, such as narrow momentum boundary layers, abrupt multiphase interfaces, and tightly concentrated charge distributions. These localized topological features strictly determine critical physical mechanisms including interfacial transport resistance, cross-field energy coupling, and the onset of localized electrohydrodynamic instabilities. Therefore, a physically robust computational model must rigorously retain the precise topology, spatial location, and field intensity of these extreme localized features while concurrently maintaining the globally consistent evolution of the coupled hydrodynamic fields.

This strict hydrodynamic requirement is explicitly addressed by the specific physical phenomena isolated within the three evaluated benchmark configurations. In the first configuration, the primary computational challenge involves the amplified near-electrode boundary layer, where steep spatial gradients directly dictate the localized electrohydrodynamic forcing and subsequent near-wall convective transport. The second configuration introduces the challenge of the radially abrupt multiphase interface, which is highly representative of complex multiphase systems where annular transitions strictly govern the spatial extent and intensity of cross-field energy exchange. Finally, the third configuration isolates a compact charged-core structure governed by sharp radial switching, representing scenarios where highly concentrated electrostatic information dominates the primary fluid transport pathways and overall flow regulation. Collectively, these configurations represent the most notoriously difficult forms of localized structural complexity frequently encountered in advanced computational fluid dynamics.

Extrapolating from these comprehensive comparative results, a broader conclusion regarding physics-informed computational modeling can be established. A predictive framework that demonstrates robust physical fidelity across these structurally diverse hydrodynamic scenarios is intrinsically far more reliable for investigating realistic multiscale applications involving localized electrostatic forcing, concentrated thermal convection, and charge-driven flow regulation. In this rigorous physical context, the RA-PINN architecture provides a highly meaningful computational advancement. The integrated residual learning mechanism significantly improves optimization stability to support coherent global flow propagation, while the parallel attention gating drastically enhances numerical sensitivity to the most dynamically informative local flow regions. This specific structural combination successfully satisfies the dual fluid dynamics requirement of preserving broad global conservation laws while achieving high-fidelity localized gradient resolution, proving particularly advantageous for advanced flow control analysis and the physics-guided simulation of complex multiscale fluid systems.

\section{Conclusions}
\label{sec:conclusion}

This study introduces a Residual-Attention Physics-Informed Neural Network designed to explicitly overcome the severe numerical diffusion inherently suffered by conventional surrogate models when predicting complex electro-thermoconvective flows. The predictive capability of the proposed architecture was rigorously evaluated across three highly localized hydrodynamic configurations, specifically encompassing an exponential electro-convective boundary layer, a radially abrupt multiphase interface, and a highly compact charged core. Quantitative and qualitative comparisons definitively demonstrate that the RA-PINN significantly outperforms both standard pure PINN and recurrent LSTM-PINN baselines in accurately resolving these steep spatial gradients. By adaptively modulating feature propagation through targeted attention gating, the network successfully preserves critical local topological fidelity without sacrificing the broader physical consistency strictly dictated by the governing multi-physics equations. Ultimately, this methodology establishes a highly robust computational framework for faithfully reconstructing extreme interfacial and boundary-layer phenomena, providing a highly valuable analytical tool for the advanced simulation of multiscale fluid systems.

\section*{Declaration of competing interest}
The authors declare that they have no known competing financial interests or personal relationships that could have appeared to influence the work reported in this paper.

\section*{Acknowledgment}
This work is supported by the Developing Project of Science and Technology of Jilin Province (20250102032JC).

\section*{Data availability}
All the code for this article is available open access at a Github repository available at https://github.com/Uderwood-TZ/RA-PINN-for-Sharp-Interfaces-and-Localized-Charge-Structures.git.

\clearpage
\onecolumngrid
\appendix
\section{Implementation settings extracted from the released scripts}

This appendix summarizes the main implementation settings used in the released scripts for Cases~1--3. The settings below are extracted from the Python files in the provided code package and are included here to improve reproducibility.

\subsection{Main training and architecture settings}

\begin{table}[htbp]
\caption{Comprehensive summary of the computational implementation settings, network architectures, and physical parameters utilized in the physics-informed training for Cases 1--3. This detailed configuration ensures the reproducibility of the reported electro-thermoconvective flow simulations.}
\label{tab:appendix_impl_settings}
\centering
\begin{ruledtabular}
\begin{tabular}{lccc}
\textbf{Numerical Setting} & \textbf{pure PINN} & \textbf{LSTM-PINN} & \textbf{RA-PINN} \\
\hline
Random seed & 20260308 & 20260308 & 20260308 \\
Numerical precision & float64 & float64 & float64 \\
Input / Output dimension & 2 / 5 & 2 / 5 & 2 / 5 \\
Hidden layer dimension & 128 & 128 & 128 \\
Backbone depth / block count & 6-layer MLP & 2 LSTM layers & 6 RA blocks \\
Feature branch mapping & Linear--Tanh & LSTM propagation & Linear--Tanh--Linear \\
Gating / Recurrent mechanism & None & Single-layer LSTM & Gated Sigmoid \\
Output head architecture & Linear & MLP head & MLP head \\
Normalization layer & None & None & LayerNorm \\
Total training iterations & 50,000 & 50,000 & 50,000 \\
Optimizer & Adam & Adam & Adam \\
Initial / Minimum learning rate & $8.0\times10^{-4} / 1.0\times10^{-5}$ & $8.0\times10^{-4} / 1.0\times10^{-5}$ & $8.0\times10^{-4} / 1.0\times10^{-5}$ \\
Weight decay & $1.0\times10^{-10}$ & $1.0\times10^{-10}$ & $1.0\times10^{-10}$ \\
Learning-rate scheduler & CosineAnnealing & CosineAnnealing & CosineAnnealing \\
Supervised grid size & $91\times91$ & $91\times91$ & $91\times91$ \\
Interior collocation points ($N_r$) & 16,000 & 16,000 & 16,000 \\
Boundary points per edge & 360 & 360 & 360 \\
Data / Collocation batch size & 1024 / 2048 & 1024 / 2048 & 1024 / 2048 \\
PDE loss weight ($\mathcal{L}_{\mathrm{pde}}$) & 1.0 & 1.0 & 1.0 \\
Data loss weight ($\mathcal{L}_{\mathrm{data}}$) & 18.0 & 18.0 & 18.0 \\
Boundary loss weight ($\mathcal{L}_{\mathrm{bc}}$) & 25.0 & 25.0 & 25.0 \\
Warm-up steps (PDE) & 5,000 & 5,000 & 5,000 \\
Residual weight (Continuity) & 2.0 & 2.0 & 2.0 \\
Residual weight (Momentum/Energy) & 1.0 & 1.0 & 1.0 \\
Spatial domain $\Omega$ & $[-1,1]^2$ & $[-1,1]^2$ & $[-1,1]^2$ \\
Viscosity coefficient $\nu$ & 0.035 & 0.035 & 0.035 \\
Thermal diffusivity $\alpha$ & 0.020 & 0.020 & 0.020 \\
Electric / Thermal force coeff. & 0.45 / 0.22 & 0.45 / 0.22 & 0.45 / 0.22 \\
Joule heating coefficient & 0.08 & 0.08 & 0.08 \\
Potential reaction coeff. $\lambda_{\phi}$ & 1.15 & 1.15 & 1.15 \\
\end{tabular}
\end{ruledtabular}
\end{table}

\subsection{Case-specific benchmark parameters}

\begin{table}[htbp]
\caption{Case-specific physical and geometric parameters for the three electrohydrodynamic benchmark scenarios. These parameters define the localized structural complexity and coupled field distributions used for the Method of Manufactured Solutions.}
\label{tab:appendix_case_parameters}
\centering
\begin{ruledtabular}
\begin{tabular}{lccc}
\textbf{Physical Parameter} & \textbf{Case 1} & \textbf{Case 2} & \textbf{Case 3} \\
\hline
Primary structural coefficient & $\beta_{\mathrm{layer}}=7.8$ & $\beta_{\mathrm{ring}}=18.0$ & $\beta_{\mathrm{step}}=45.0$ \\
Characteristic geometric scale & freq. $=1.0$ & $r_{\mathrm{in}}=0.26, r_{\mathrm{out}}=0.58$ & $r_{0}=0.5$ \\
Auxiliary radial scale & --- & $r_{\mathrm{core}}=0.24$ & $r_{0}^{2}=0.25$ \\
Core center coordinates & --- & $(0,0)$ & --- \\
Velocity-field ($\psi$) coefficients & $A=0.18, B=0.08$ & $A=0.16, B=0.09, \gamma=3.2$ & $A=0.24, B=0.10$ \\
Pressure-field coefficients & $A=0.46, B=0.18, C=0.14$ & $A=0.42, B=0.18, C=0.12$ & $A=0.75$ \\
Temperature-field coefficients & $T_0=0.42, A=0.58$ & $T_0=0.38, A=0.52$ & $T_0=0.35, A=0.55$ \\
Electric potential coefficients & $A=1.15, B=0.26$ & $A=1.05, B=0.62$ & $A=0.75, B=0.15$ \\
Decay / exponential parameters & $\beta_{\psi}, \beta_{p}, \beta_{T}, \beta_{\phi}$ series & --- & --- \\
\end{tabular}
\end{ruledtabular}
\end{table}

\clearpage
\twocolumngrid


%
%

%


\end{document}